\documentclass[twocolumn,showpacs,preprintnumbers,amsmath,amssymb]{revtex4}
\usepackage{graphicx,epsfig}
\usepackage{dcolumn}
\usepackage{bm}

\begin{document}
\topmargin .2cm
\title{DILEPTON EMISSION AT TEMPERATURE DEPENDENT BARYONIC QUARK-GLUON PLASMA}
\author{\bf S. Somorendro Singh\footnote{Email: sssingh@physics.du.ac.in }  and Yogesh Kumar}

\affiliation{ Department of Physics and Astrophysics, University of Delhi, Delhi - 110007, India}
\begin{abstract}
  A fireball of QGP is evoluted at temperature
dependent chemical potential by a statistical model in the pionic
medium. We study the dilepton emission rate at temperature dependent chemical 
potential (TDCP) from such a fireball of QGP. In this model, we take 
the dynamical quark mass as 
a finite value dependence on temparature and  parametrization factor of
the QGP evolution. The temperature and factor in quark mass 
enhance in the growth of the droplets as well
as in the dilepton emission rates. 
The emission rate from   
the plasma shows dilepton spectrum 
in the intermediate mass region (IMR) of$~ (1.0-4.0)$~GeV and its
rate is observed to be a strong increasing function of 
the temperature dependent chemical potential for quark and 
antiquark annihilation.     
\end {abstract} 
\maketitle
\section{\bf Introduction}  
 \par  The
ongoing experiments like ultra-relativistic heavy-ion
collision at BNL and the large hadron collider at CERN have 
focussed on the search of 
the QCD phase structure and the formation of mini big bang. 
The experiments at BNL and
CERN will provide the best platform to study 
the creation and evolution of such mini big bang called 
Quark-Gluon Plasma (QGP), which 
is perhaps believed
to be formed in the expansion of the early universe~\cite{karch}. 
Since we believe that 
the matter existed
only for a few microseconds after the big-bang, 
its direct detection is very difficult even in these experiments. There
are indirect possibilities for detection like strangeness enhancement~\cite{rafelski}, $J/\psi$
suppression~\cite{matsui} and radiation of 
dileptons and photons~\cite{shuryak,kajantie,srivastava} etc.
Among these indirect probes,
dileptons and photons are considered to be the most promising
signals for its detection of QGP formation created in relativistic 
heavy-ion collision (RHIC). It is due to the fact that the 
dilepton driving out of the collisions among the quarks, antiquarks and 
gluons brings the whole informations about the existence of the plasma
fireball and tell the properties of the fireball to the detector.
They interact through electromagnetic force due to the 
large mean free path in their production. In order to see the
production of dilepton for 
the signal of QGP formation, we look
at the process of annihilation of quark and antiquark and they produce 
virtual 
photons which subsequently decay 
into dileptons 
such as $l^{-}l^{+},~\mu^{+}\mu^{-}$.
\par
  Many theoretical and experimental researchers have 
calculated dilepton and photon emissions 
at finite temperature and at quark chemical potential. 
The experiments at AGS 
and SPS energies~\cite{nagamiya} have reported the presence of significant 
amount of baryon chemical potential and 
even  at RHIC energies~ $\sqrt{s} \leq 200\, A GeV$ there has been
the detection of such baryon chemical potential. 
These informations indicate 
~\cite{gustafson,mohring} that the colliding heavy ions may not 
be fully transparent in the centrality region of the colliding 
particles and the region may have significant amount of
dense nuclear matter. The work of Hammon and coworkers~\cite{hammon} 
supported these arguments of existing the chemical potential and predicted
the initial nonequilibrium QGP produced at RHIC energies, indicating that
the system has finite baryon density or chemical potential. 
So, in the theoretical study, dilepton emissions in finite baryonic chemical potential has been calculated
through various distribution functions and perhaps, the 
work of Dumitru et al.~\cite{dumitru} gives the first signal
of dilepton emission at finite baryonic chemical potential with
Fermi distribution functions. Then this work is further studied 
by Strickland using
quark and gluon fugacities in j$\ddot{u}$ttner distribution function. 
 He
showed another promising result calculated
from the non-equilibrium
quark-gluon plasma ~\cite{strickland}.
 The recent work of Majumder et al.~\cite{majumder} have indicated
the emission of 
dileptons from QGP at the RHIC energies at 
finite baryon density. Bass et al. ~\cite{bass} idea of   
parton rescattering  and fragmentation leads to  a substantial 
increase in the net-baryon density 
at midrapidity region. Besides these works, we have the reports of 
other authors on dilepton production at low mass region~\cite{rapp}.  
These works suggest the importance of chemical 
potential in the dilepton calculation. To produce
such emission, we consider the system in the pionic medium in which 
the equilibrium thermodynamic QGP is
a function of temperature~ $T$ and chemical potential$~\mu~$ and the
potential itself as function of temperature . 
\par In this brief article, we choose the baryonic chemical potential which
is considered to be temperature dependent chemical potential (TDCP) 
and 
the value has a change on the quark and antiquark distribution functions. 
We take the value of the chemical potential
in the scale of QCD parameter of dense nuclear matter. The 
chemical potential considered is obtained through~\cite{hamieh}:
\begin{equation}
   \mu(T)=2\pi\beta^{-1}\sqrt(1+\frac{1}{\pi^{2}}\ln^{2}\lambda_{q})
\end{equation}
where $T=\frac{1}{\beta}$ taken in the scale of QCD and $\lambda_{q}=e^{\frac{\mu_{q}}{T}}~$is
quark fugacity.
However, we consider 
the massless dynamical quark as a finite 
value and it is called thermal dependent quark mass (TDQM) 
obtained through the parametrization value and temperature. The finite 
value of 
the quark mass is defined as
\begin{equation}  
m^{2}_{q}= \frac{8 \pi}{(33-3n_{f})}\frac{T^{2}}{\ln(1+(\frac{\gamma N^{\frac{1}{3}} T^{2}}{2 \Lambda^{2}})^{\frac{1}{2}})} 
\end{equation}
with the QCD parameter~ $\Lambda = 150~MeV$ 
and normalising $~ N=\frac{16 \pi}{27}$. $ \gamma $ is parametrization 
factor which is like the Reynold's number to take care of 
the hydrodynamical aspects of the hot QGP flow. Its value is determined
in the most effective way of the flow parameter of quarks~ $\gamma_{q}$ 
and gluons~ $\gamma_{g}$ and it enhances in the growth of
free energy of QGP droplet.
It is expressed as
\begin{equation}
\gamma=\sqrt 2( \frac{1}{\gamma_{q}^2}+\frac{1}{\gamma_{g}^2})
\end{equation}
with the value of~ $\gamma_{g}= 6 \gamma_{q}~~ or~~ 8 \gamma_{q}$~ 
and~
$\gamma_{q}=1/6$~\cite{yen}.
\par
 Using all such parameters, we create the QGP droplet at TDCP 
incooperating the quark mass as a finite value in the system and look 
at the improvement produced by the TDCP in the growth of QGP droplet.
Then we calculate dilepton emission at the TDCP from such a system of
QGP and see its emission rate.
\par The paper is organised as follows: In Sec.II we recall a
 brief highlight of the 
evolution/growth of QGP fireball through the statistical model in 
the pionic medium.
In Sec.III we look at the dilepton emission and intgrated yields 
at temperature dependent 
chemical potential (TDCP). In the last Sec.IV 
we conclude and present our results. 

\par
\section{\bf   The free energy growth of QGP droplet at TDCP}
 \par  We use the statistical model at the TDCP for 
the growth of QGP droplet in the pionic medium. The system 
is considered
to be constituted by the free quarks, antiquarks, gluons and pions. 
We construct 
the free energies of the particles using the model of mean 
field potential in their density of state.
The constructed
free energies of these  
noninteracting fermions and bosons at the temperature dependent 
chemical potential are defined as~$F_{i}$ 
in which contribution of quarks and gluons are 
indicated by the upper sign or bosons by the lower sign. 
It is expressed as:
\begin{equation}
       F_{i}=\mp T g_{i}\int dp \rho(p_{i})ln(1\pm \exp(-\sqrt{m_{i}^{2}+p_{i}^{2}}+\mu/T))
\end{equation}
where  $g_{i}$ is the appropriate colour and particle-antiparticle
degeneracy factor. Its value is taken as $6$ and $8$ for quarks and
gluons~\cite{neergaard}. $\rho(p_{i})$ 
is the density of states of the 
particular particle~$i$~
(quarks,~gluons) based on the effective potential
among the interacting particles and it is defined within the range of 
momentum space $dp_{i}$ in a 
spherically symmetric
system. It is given by~\cite{ramanathan}
\begin{equation}
\rho(p_{i})=v/\pi^{2}[(-V_{eff}(p))^{2}(-\frac{dV_{eff}(p)}{dp})]
\end{equation}
where, $V_{eff} =(1/2p)\gamma g^{2}(p)T^{2}-m^{2}_{0}/2p$, known as 
mean field effective
potential among the quarks, antiquarks and quarks-gluons. 
$ g^{2}(p)$ is the 
first order QCD running coupling constant. It is given by
\begin{equation}
 g^{2}(p)=(4/3)(12\pi/27)[1/\ln(1+p^2/\Lambda^2)]
\end{equation}  
The free energy of pion contributed to the total free 
energies of quarks and gluons is given as:
\begin{equation}
       F_{\pi}=(3T/2\pi^{2}) v \int_{0}^{\infty} p_{\pi}^{2}dpln(1- \exp(-\sqrt{m_{\pi}^{2}+p_{\pi}^{2}}/T))
\end{equation}
The contribution of the pion energy is due to the fact that
the transformation of the phase is slightly dominanted by the pions over the
other hadronic particles. In addition to these free energies there is
another contribution to the total free energy and it takes the role of
bag constant in confining the system. It is called the interfacial energy given by~\cite{weyl}
\begin{equation}
     F_{interface}=\frac{1}{4}R^{2}T^{3}\gamma
\end{equation}.
Therefore, the total energy of QGP fireball is:
\begin{equation}
     F=\sum_{j} F_{j}
\end{equation}
where j stands for the different particles viz quark, gluon, pion
and interface. The free energy can indicate the nature of QGP fireball 
evolution and its transition. It also
explains the 
creation of the plasma formation with the size of the droplets.
\par
\section{\bf  Dilepton emission at TDCP from QGP}
\par  
The calculation of dilepton emission at finite temperature
and at finite baryon chemical potential have been done by many authors.
These calculations are performed on the basis of the 
expected results coming out from the heavy-ion collision experiments.
The experiments expect more productions of lepton than of other 
particles produced.
The possible sources of dilepton are from the annihilation 
 $q\bar{q}\rightarrow l^{+}l^{-}$, compton like scattering,
$q(\bar{q})g\rightarrow q(\bar{q})l^{+}l^{-}$ 
and $gg\rightarrow q\bar{q}l^{+}l^{-} $fusion processes. 
Among the processes, Drell-Yan reaction is mostly used 
for thermal emission of dilepton pairs~\cite{ruuskanen} 
and the compton scattering such as
$q(\bar{q})g \rightarrow q(\bar{q})+ l^{+}l^{-}$ follows after 
Drell-Yan reaction. In this article, we 
exclusively engage in quark-antiquark annihilation such as
 $q\bar{q}\rightarrow l^{+}l^{-}$  reaction 
for the dilepton
emission. This is due to fact that it produces larger amount of lepton
pair in comparison to the other collisions. In the process, we consider
only the dominant production of dilepton in the intermediate mass region
neglecting the dilepton spectra 
from the low mass region. This is 
fact that the contribution of dilepton through the deacy of mesons in the 
system is negligence.  
So the dilepton emission rate produced $\frac{dN}{d^{4}x}$ is given 
by~\cite{gale}:
\begin{eqnarray}
\frac{dN}{d^{4}x}&=&\int \frac{d^{3}p_{1}}{(2 \pi)^{3}} 
\frac{d^{3}p_{2}}{(2 \pi)^{3}} n_{q}(p_{1},\mu) 
n_{\bar{q}}(p_{2},\mu)\nonumber 
\\
&\times &v_{q\bar{q}} \sigma_{q\bar{q}}(M^{2}) 
\end{eqnarray}
where, 
\begin{equation}
 n_{q}(p_{1},\mu)=\frac{\lambda_{q}}{\exp^\frac{(p_{1}-\mu)}{T}+\lambda_{q}},
n_{\bar{q}}(p_{2},\mu)=\frac{\lambda_{\bar{q}}}{\exp^\frac{(p_{2}+\mu)}{T}+\lambda_{\bar q}}
\end{equation}
are Fermi-Dirac distribution functions for 
quarks and antiquarks~\cite{biro}
with their corresponding parton fugacities~$\lambda_{q(\bar{q})}=e^{\frac{\mu}{T}}$. 
For gluon, the Bose-Einstein distribution function is:
\begin{equation}
n_{g}(p,\mu)=\frac{\lambda_{g}}{\exp^\frac{p_{g}}{T}-\lambda_{g}}
\end{equation}
with parton gluon fugacity $\lambda_{g}$. The function for gluon
can be used in the collision of $qg\rightarrow  l^{+}l^{-}$ or $gg$ 
fusion reaction.  
$v_{q\bar{q}}$ is the relative velocity of annihilating quark 
pair and  $p_{\mu}$ is lepton pair four momentum.($M^{2}=p^{\mu}p_{\mu}$ 
invariant lepton pair mass ). $\sigma_{{q\bar{q}}
{\rightarrow}{l\bar{l}}}$ is the electromagnetic 
annihilation cross section. Substituting 
the distribution functions for quark
and antiquark in the  equation  $(10)$ using $(11)$, and integrate 
over $q$ and $\bar{q}$ momentum, we obtain dilepton emission rate at TDCP as:
\begin{eqnarray}
 \frac{dN}{dM^{2}d^{4}x} & = & \frac{5 \alpha^{2}}{18 \pi^{3}}
T M  e^{4 \sqrt{\pi^{2}+\ln^{2}\lambda_{q}}} \nonumber \\
                         &\times &(1+\frac{2 m_{q}^2}{M^2}) K_{1}(M/T) 
\end{eqnarray}

In the above solution, $ K_{1}(M/T) = G(z) $ which is known as the modified 
Bessel's function and
volume element is $d^4x =d^2x_{T}dy\tau d\tau$.
We expand longitudinally the above expression and finally we have emission
rate as:
\begin{eqnarray}
 \frac{dN}{dM^{2}dy} & = & \frac{5 \alpha^{2} R^2}{18 \pi^{2}}
M(1+\frac{2 m_{q}^2}{M^2})  \nonumber \\
     &\times &\int e^{4 \sqrt{\pi^{2}+\ln^{2}\lambda_{q}}/T(\tau)} G(z,\tau)T(\tau) \tau d\tau
\end{eqnarray}
where,~ $ T(\tau)= T_{0}(\frac{\tau_{0}}{\tau})^{1/3}$.

\section{\bf  Results and Conclusions}
\par The results of evolution of QGP
fireball are
shown in the figures~$(1-3)$. Figure~$(1)$ shows the evolution
of free energy at zero chemical potential at the parametrization value
of $~\gamma_{q}=1/6,~\gamma_{g}= 6 \gamma_{q}$. It shows
very much stability in the formation of droplets for the various
values of temperature. Figure~$(2)$ shows the change of evolution
of the free energy for these various
values of chemical potential and indicate the decreasing size of droplets
with the increasing chemical potential. The figure~$(3)$ shows
the change of free energy 
with the size of droplet at temperature
dependent chemical potential with the finite value of quark chemical 
potential$~\mu_{q}= 47 MeV$. The result indicates that the evolution of
QGP through the model has first order shift at the 
temperature~$T=(150-170)~MeV$ with the increase of chemical potential
and the size of the droplet is increased with the increase of the chemical
potential. It implies that the evolution of QGP droplet is enhanced
with the effect of temperature dependent chemical potential.
Moreover the shift in the first order is further explained 
in figure~$(4)$ by the behaviour of the entropy v/s chemical potential. 
There is a clear jump
in the continuity of the entropy curve at the
chemical potential~$\mu=(990-1005)~MeV$ at which we found the corresponding 
temperature as~$T=(150-170)~MeV$.
\par  In figure~$(5)$, we show
dilepton emission for various values of 
initial
temperature~$T_{0}$
and at transition temperature~$T_{c}=0.17~GeV$ without the chemical
potential and compared the results with other theoretical calculations
of dilepton emission at $~\mu=0$. The results are same over the range
of lepton pair mass $~M~$. In the figure~$(6)$, we show the comparison of 
emission rates of dilepton at 
the temperature
dependent chemical potential and at finite chemical potential. 
The emission rate increases with the increase
of temperature dependent chemical potential at 
the transition temperature~$T_{c}=0.17~GeV$ over the finite chemical potential.
The emission rate is much higher
at the temperature dependent chemical potential than the emission at finite
chemical potential. 
\par Now, if we look dilepton yields with the change of the lepton pair mass 
in above figures of dilepton production, we obtain
a sudden fall in the production rate of dilepton with increase in
lepton pair mass~$M$ upto~$3~GeV$. It indicates that at higher 
lepton pair mass more suppression is obtained in comparison to the 
low mass region. 
\par
We look again at dilepton integrated yields with the evolution time of the 
QGP droplet. The dilepton integrated yield is exponentially
increasing with the evolution time and after a certain time, it becomes
constant for these different values of chemical potential. The
plots are shown in the figures~$(7-8)$ for the chemical potential
with their corresponding initial
temperatures at transition temperature~$T=0.17~GeV~$. 
Figure~$(7)$ shows the integrated yields at transition
temperature $T_{c}=0.17~GeV$ without the finite value of chemical 
potential and again compared the results with other results. They are same
at this zero chemical potential.  
 In figure~$(8)$, we plot the
integrated yield for the temperature dependent chemical potential and
finite chemical potential at
the transition temperature~$T=0.17~GeV$. The integrated yield is
very large as compared to the results of the earlier work of integrated
yeild at finite value of chemical potential. This is due to the
fact that both temperature and chemical potential enhance in the 
interaction of the particles of the system and yields more dileptons.
\par
Now it indicates that the emission and integrated yield are higher at TDCP 
as compared to the subsequently fixed value of chemical potential and at
zero chemical potential. 
It means in the QGP phase where the 
temperature and chemical potential coexist together then the interacions
among 
the partciles are more and dileptons are produced more. 
So the model with the temperature dependent chemical potential produced
improvement over results of finite baryonic chemical potential and zero
chemical potential.

\begin{figure}
\resizebox*{3.0in}{3.0in}{\rotatebox{360}{\includegraphics{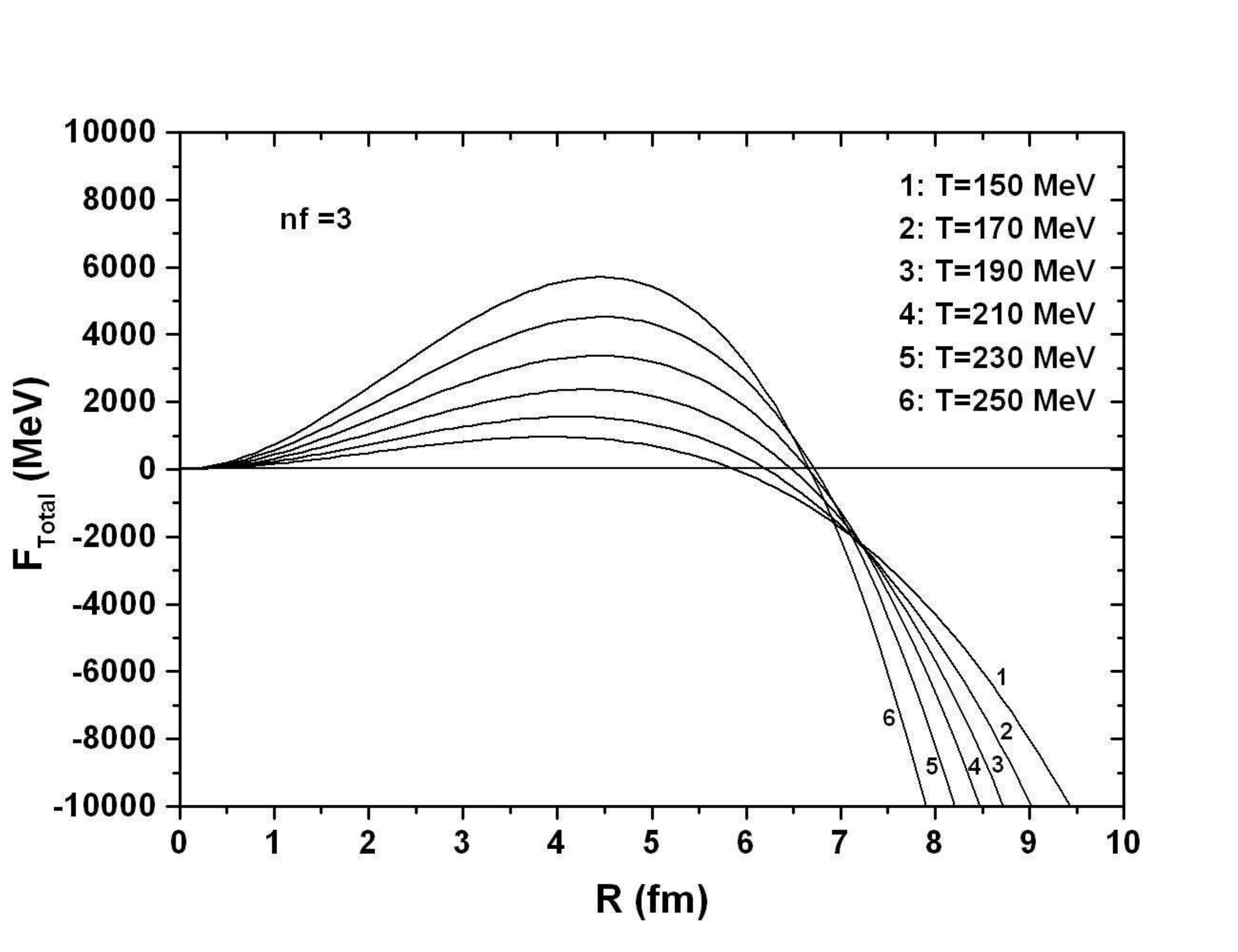}}}
\vspace*{0.5cm}
\caption[]{
Free energy with the change of droplet size for various values of
temperature.}
\label{scaling}
\end{figure}

\begin{figure}
\resizebox*{3.0in}{3.0in}{\rotatebox{360}{\includegraphics{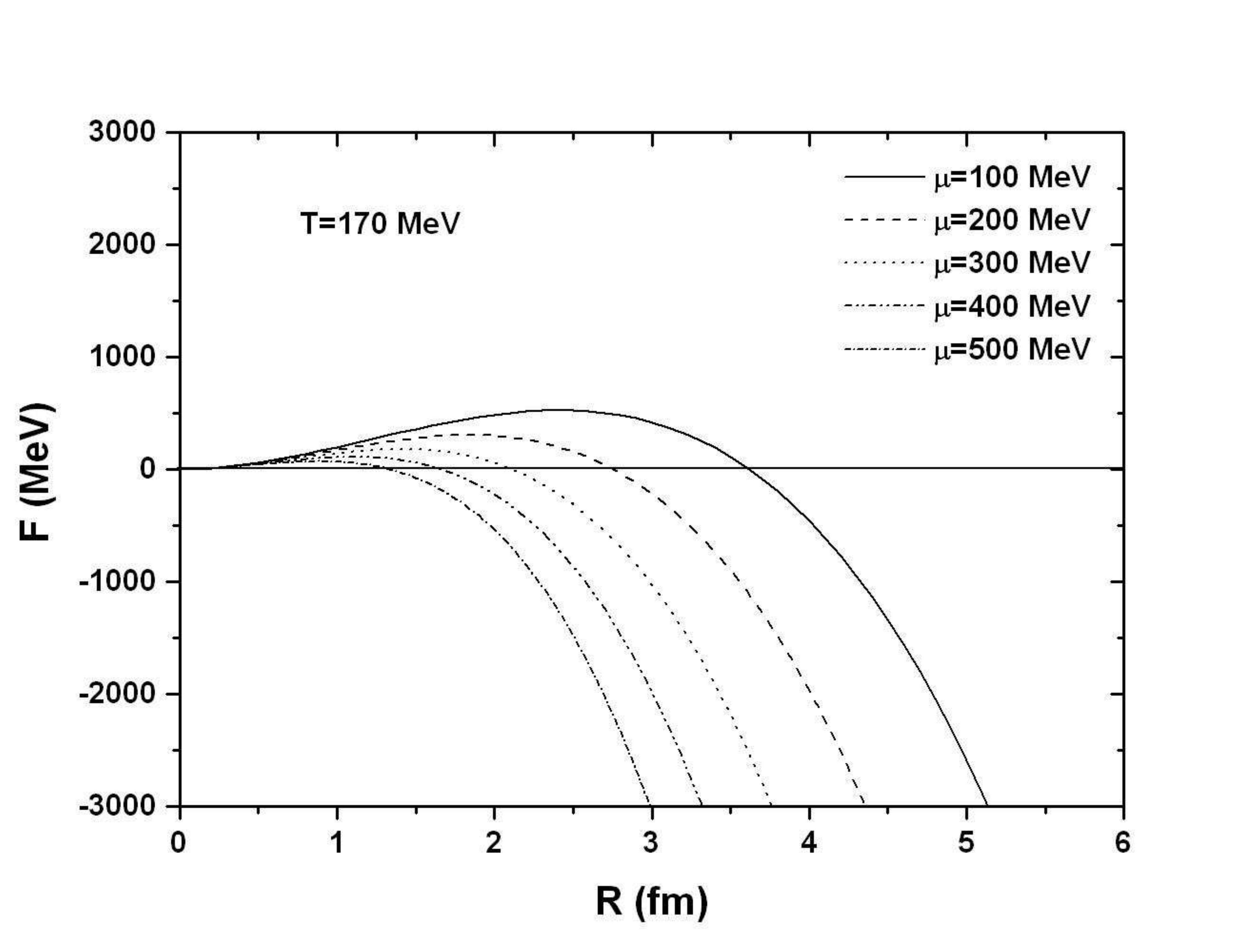}}}
\vspace*{0.5cm}
\caption[]{
Free energy with the change of droplet size at the particlur 
transition temperature~$ T_{c}=0.17~GeV $ for variuos values of
chemical potential.}
\label{scaling}
\end{figure}

\begin{figure}
\resizebox*{3.0in}{3.0in}{\rotatebox{360}{\includegraphics{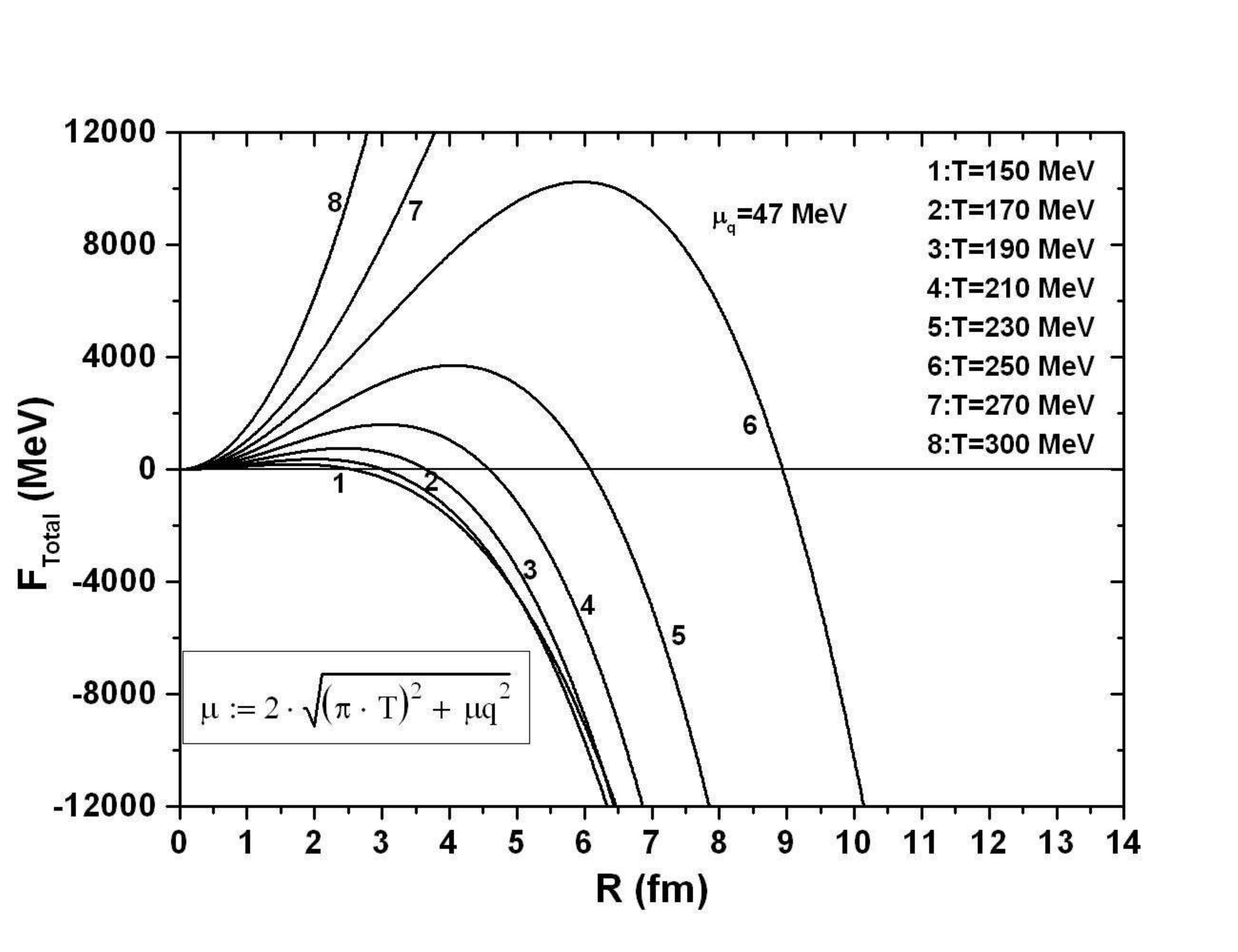}}}
\vspace*{0.5cm}
\caption[]{
Free energy with the change of droplet size 
at the particular quark chemical potential~$\mu_{q}=0.47~GeV$
for different initial temperatures.}
\label{scaling}
\end{figure}

\begin{figure}
\resizebox*{3.0in}{3.0in}{\rotatebox{360}{\includegraphics{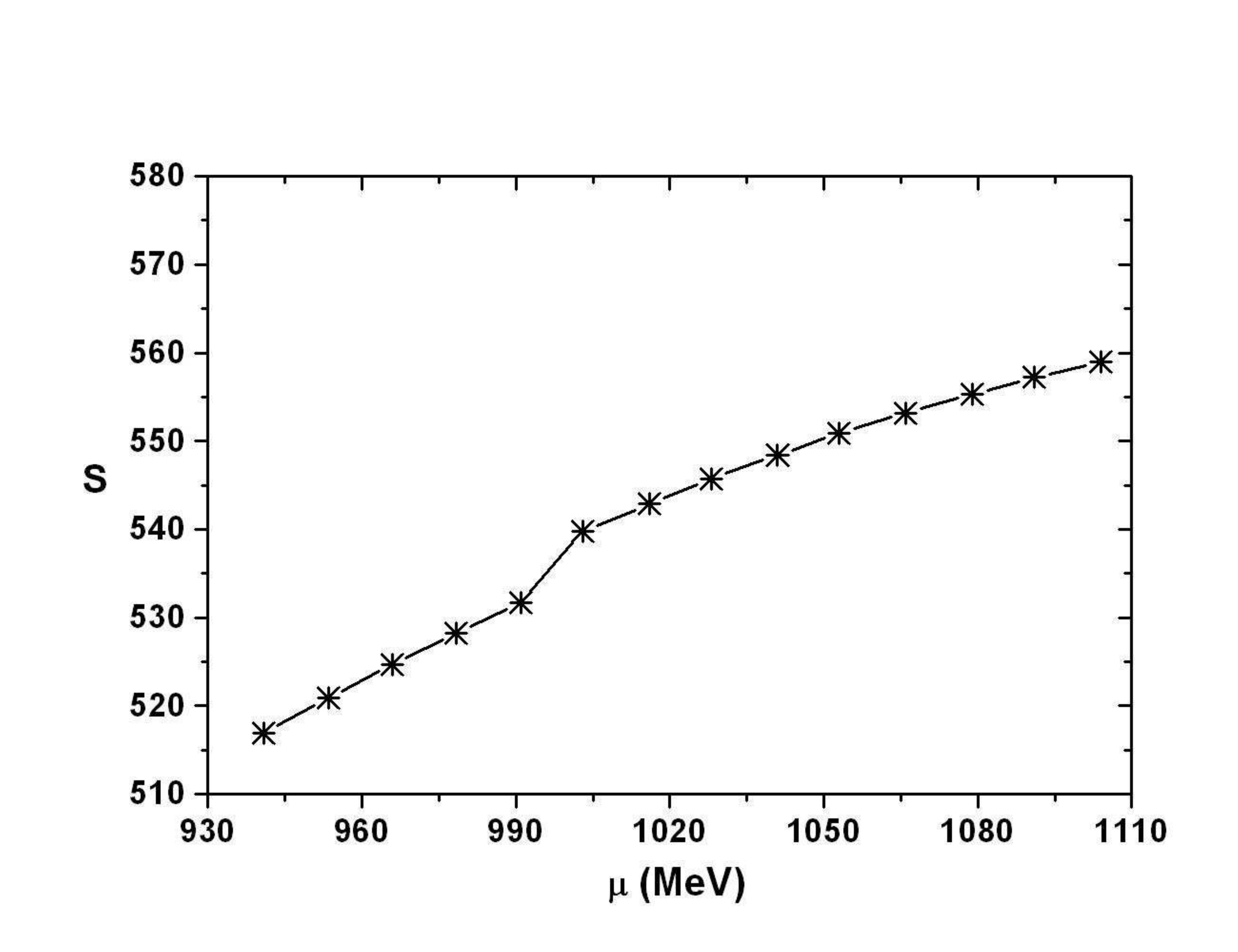}}}
\vspace*{0.5cm}
\caption[]{
Entropy v/s, change of chemical potential~$\mu $ and first order transition
at the temperature~$T_{c}=
0.17~GeV$.}
\label{scaling}
\end{figure}
\begin{figure}
\resizebox*{3.0in}{3.0in}{\rotatebox{360}{\includegraphics{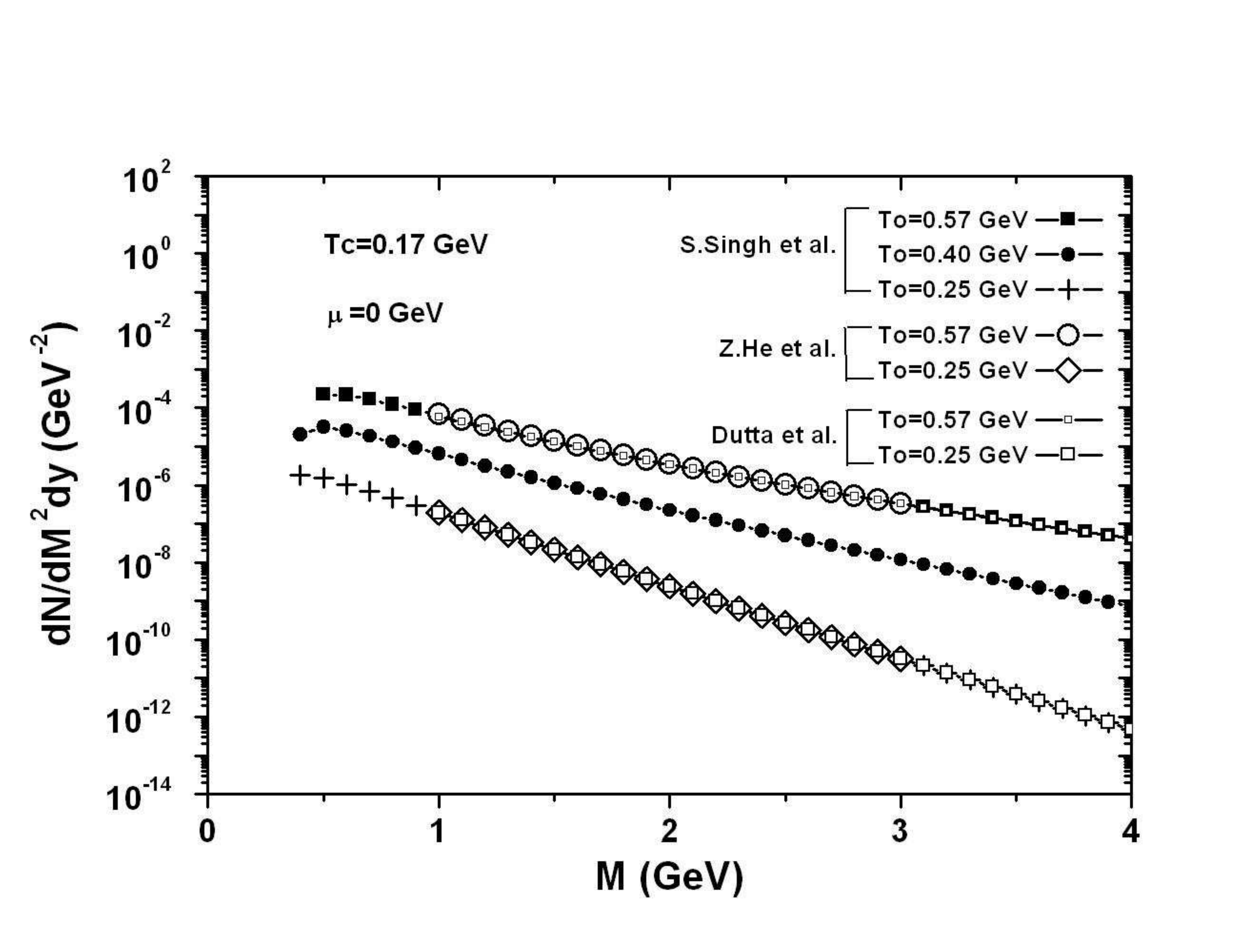}}}
\vspace*{0.5cm}
\caption[]{
The dilepton emission rate,~$\frac{dN}{dM^{2} dy}~(GeV^{-2}) $, 
at transition temperature~$T_{c} 
=0.17~GeV$ and at zero chemical potential for different initial temperatures
and its compared curve. 
}
\label{scaling}
\end{figure}
\begin{figure}
\resizebox*{3.0in}{3.0in}{\rotatebox{360}{\includegraphics{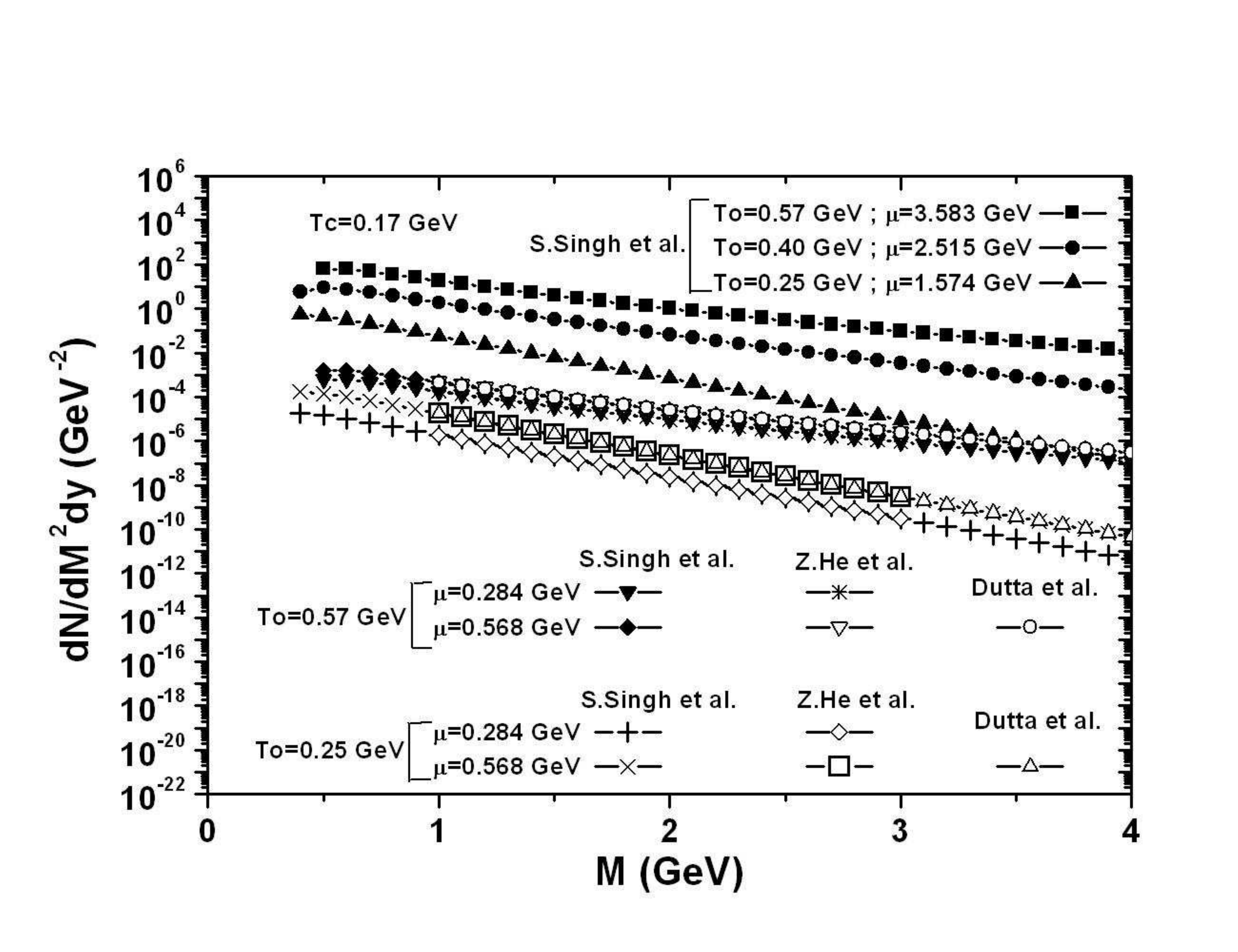}}}
\vspace*{0.5cm}
\caption[]{
The dilepton emission rate,~$\frac{dN}{dM^2 dy}~(GeV^{-2}) $,
 at transition temperature~$T_{c}
=0.17~GeV$ for the different values of
~$\mu $ with different initial temperatures and its compared curves.}
\label{scaling}
\end{figure}

\begin{figure}
\resizebox*{3.0in}{3.0in}{\rotatebox{360}{\includegraphics{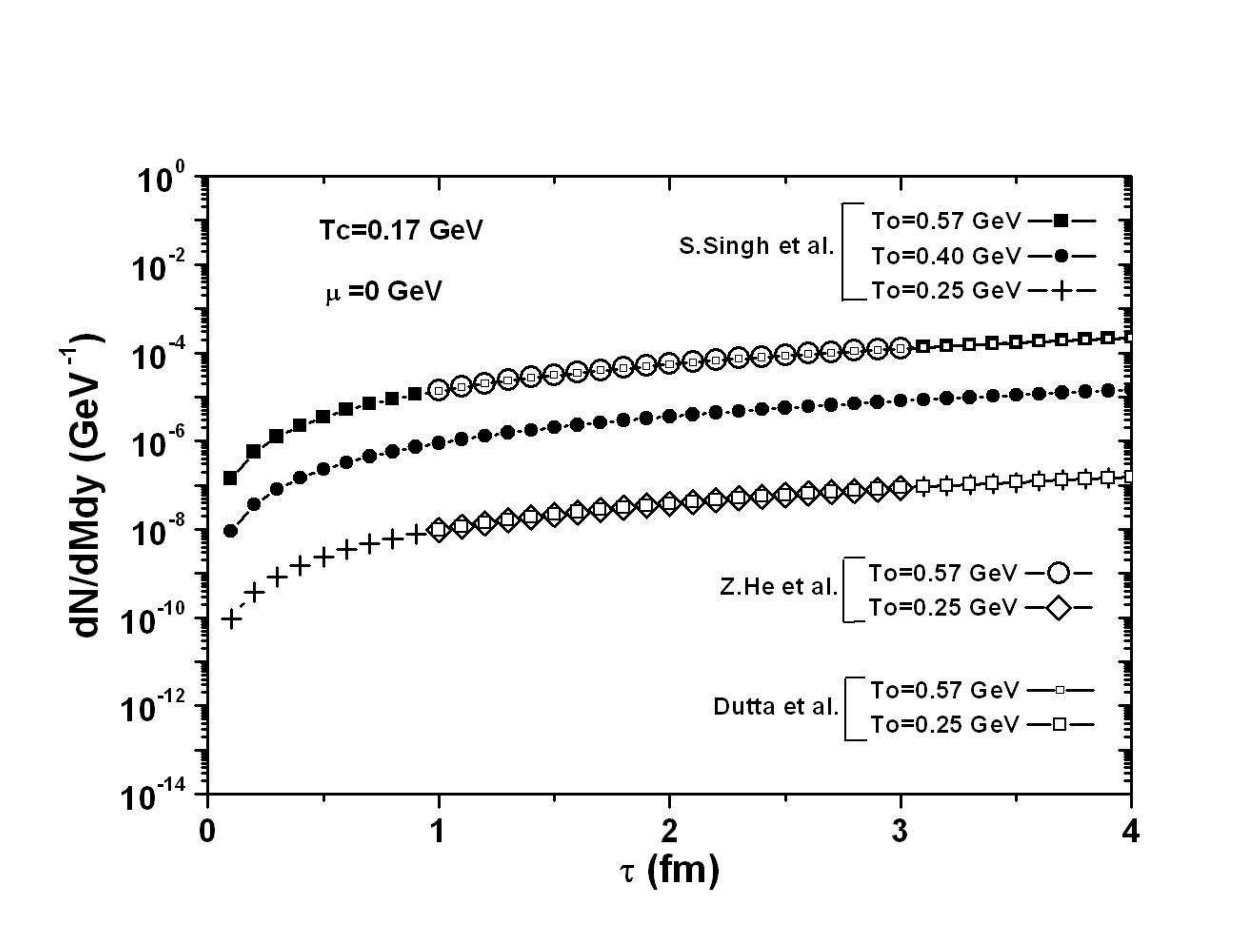}}}
\vspace*{0.5cm}
\caption[]{
The dilepton integrated yields,~$\frac{dN}{dM dy}~(GeV^{-1}) $,
 at transition temperature~$T_{c}
=0.17~GeV$ and at zero chemical potential 
~$(\mu=0) $ with different initial temperatures and its compared curves.}
\label{scaling}
\end{figure}

\begin{figure}
\resizebox*{3.0in}{3.0in}{\rotatebox{360}{\includegraphics{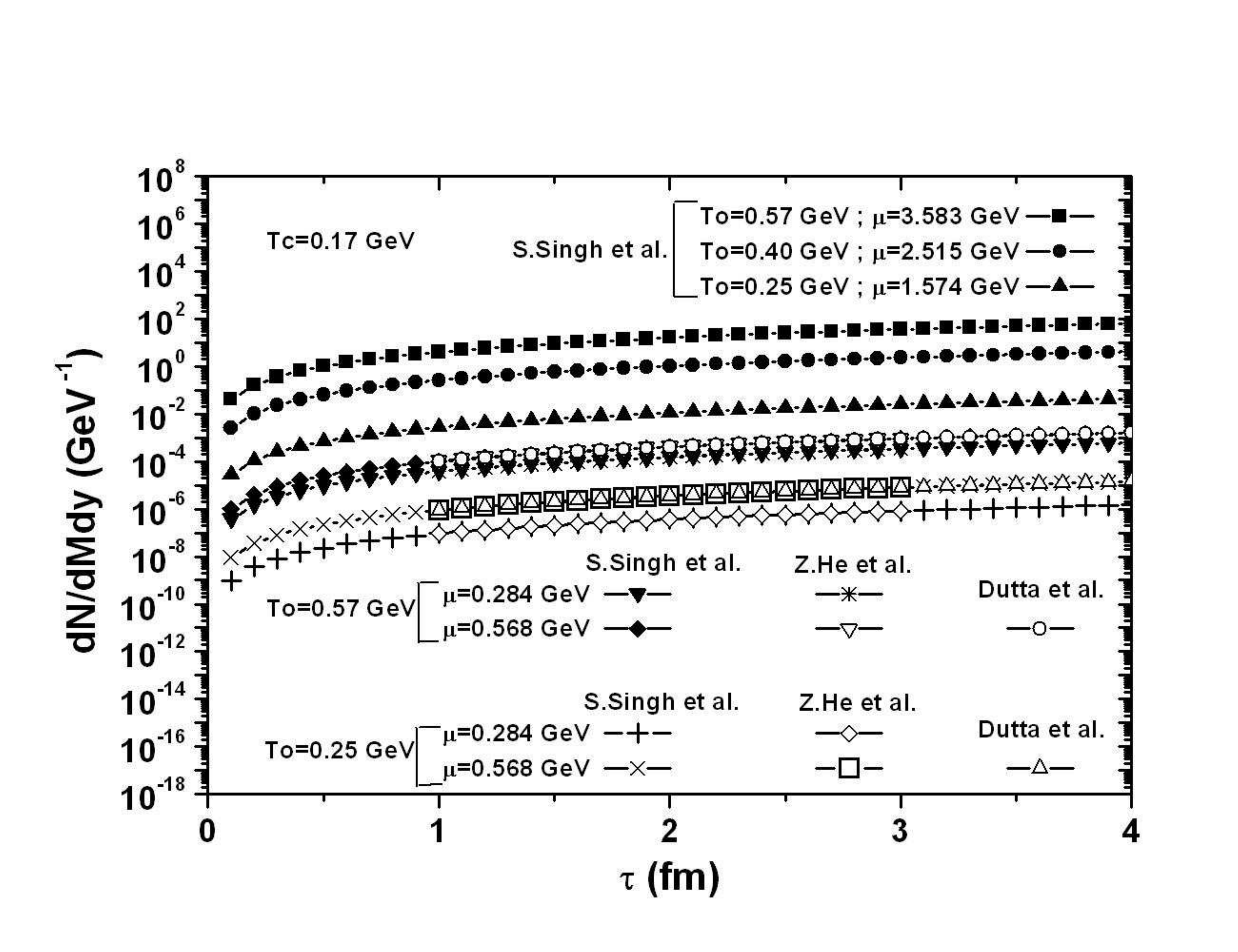}}}
\vspace*{0.5cm}
\caption[]{
The dilepton integrated yield,~$\frac{dN}{dM dy}~(GeV^{-1}) $,
 at transition temperature~$T_{c}
=0.17~GeV$ for the different values of
~$\mu $ with different initial temperatures and its compared curves.}
\label{scaling}
\end{figure}
\acknowledgements
We are very thankful to Dr. R. Ramanathan for his 
constructive 
suggestions and discussions. The author (YK) would like to thank the
Department for research facility and highly oblized to express his gratitute 
to Rajiv Gandhi Fellowship, UGC, New Delhi for the financial support.

\end{document}